\documentclass[aps,prl,twocolumn,superscriptaddress,english,amsmath,amssymb]{revtex4-2}

\usepackage{graphicx}
\usepackage{dcolumn}
\usepackage{braket}
\usepackage{amsmath} 
\usepackage{amsmath, amssymb, amsfonts}
\usepackage{color}

\usepackage{array}

\usepackage{esint}

\usepackage{bm}

\usepackage{bbm}
\usepackage[breaklinks,colorlinks,
linkcolor=blue,citecolor=blue,urlcolor=blue]{hyperref}
\usepackage{graphicx,amssymb, color}
\usepackage{mathtools}
\usepackage{soul}
\usepackage[dvipsnames]{xcolor}
\usepackage{tikz}
\usepackage{tcolorbox}
\usepackage{graphics,amssymb,amsmath,epsfig,color}
\usepackage{physics}
\usepackage{orcidlink}
\usepackage{ulem}

\begin{document}

\preprint{APS/123-QED}

\title{Observation of a supersolid stripe state in two-dimensional dipolar gases}

\author{Yifei He}
\affiliation{Department of Physics, The Hong Kong University of Science and Technology,
Clear Water Bay, Kowloon, Hong Kong}%

\author{Haoting Zhen}
\affiliation{Department of Physics, The Hong Kong University of Science and Technology,
Clear Water Bay, Kowloon, Hong Kong}%

\author{Mithilesh K. Parit}
\affiliation{Department of Physics, The Hong Kong University of Science and Technology,
Clear Water Bay, Kowloon, Hong Kong}%

\author{Mingchen Huang}
\affiliation{Department of Physics, The Hong Kong University of Science and Technology,
Clear Water Bay, Kowloon, Hong Kong}%

\author{Nicol\`o Defenu}
\affiliation{Institut f\"ur Theoretische Physik, ETH Z\"urich, Wolfgang-Pauli-Str. 27 Z\"urich, Switzerland.}%

\author{Jordi Boronat}
\affiliation{Departament de Física, Universitat Politècnica de Catalunya, Campus Nord B4-B5, 08034 Barcelona, Spain}%

\author{Juan Sánchez-Baena}
\affiliation{Departament de Física, Universitat Politècnica de Catalunya, Campus Nord B4-B5, 08034 Barcelona, Spain}%

\author{Gyu-Boong Jo}
\altaffiliation{email: gbjo@rice.edu}
\affiliation{Department of Physics and Astronomy, Rice University, Houston, TX, USA}%
\affiliation{Smalley-Curl Institute, Rice University, Houston, TX, USA}%
\affiliation{Department of Physics, The Hong Kong University 
of Science and Technology,
Clear Water Bay, Kowloon, Hong Kong}%

\date{\today}



\begin{abstract}
     Fluctuations typically destroy long-range order in two-dimensional (2D) systems, posing a fundamental challenge to the existence of exotic states like supersolids, which paradoxically combine solid-like structure with frictionless superfluid flow. While long-predicted, the definitive observation of a 2D supersolid has remained an outstanding experimental goal. Here, we report the observation of a supersolid stripe phase in a strongly dipolar quantum gas of erbium atoms confined to 2D. We directly image the periodic density modulation, confirming its global phase coherence through matter-wave interference and demonstrating its phase rigidity relevant to the low-energy Goldstone mode, consistent with numerical calculations. Through collective excitation measurements, we demonstrate the hydrodynamic behavior of the supersolid. This work highlights a novel mechanism for supersolid formation in low dimensions, and opens the door for future research on the intricate interplay between temperature, supersolidity, and dimensionality.

\end{abstract}

\maketitle


{\textbf{Introduction.}}
Supersolidity combines the frictionless flow of a superfluid with the crystal-like periodic density modulation of a solid. It simultaneously breaks the U(1) symmetry, giving rise to global coherence, and the translational symmetry, resulting into crystalline order. This long-predicted phase has been extensively studied in various systems in the recent years~ ranging from dipolar atoms~\cite{tanzi2019observation, bottcher2019transient, chomaz2019long, guo2019low, tanzi2019supersolid, tanzi2021evidence,norcia2021two,PhysRevLett.128.195302}, spin-orbit-coupled bosons~\cite{li2017stripe,chisholm2024probing}, atoms coupled to cavity fields~\cite{leonard2017monitoring,leonard2017supersolid} to non-equilibrium atomic~\cite{liebster2025supersolid} and polariton condensates~\cite{trypogeorgos2025emerging}. Despite significant progress in understanding supersolid phases, most current realizations remain limited to regimes where dynamics are essentially three-dimensional~(3D).

 Transitioning to two dimensions (2D) presents a profound challenge. Enhanced thermal fluctuations in 2D are known to disrupt long-range order, as formalized by the Mermin-Wagner theorem~\cite{mermin1966absence}, seemingly forbidding the existence of a true, infinite crystal at finite temperatures. While a unique form of quasi-long-range order can support superfluidity through the Berezinskii-Kosterlitz-Thouless (BKT) transition~\cite{berezinskii1972destruction,kosterlitz1973ordering}, it has remained a central open question whether this phase could coexist with crystalline quasi-order. Earlier works with $^4$He in graphene layers suggest 2D superfluids in a periodic potential may reveal unconventional quantum phases beyond the BKT picture~\cite{Nyki2017,Lieu2019,gordillo_PRL:2020,Choi2021mme,knapp2025thermodynamic}, but this issue remains controversial.
 
 

 Strongly dipolar quantum gases have emerged as a promising platform to explore this frontier in 2D~\cite{zhen2025breaking}, yet creating such systems remains challenging.  While both BKT transitions in weakly dipolar 2D gases~\cite{he2025exploring} and metastable stripe crystals in strongly dipolar 2D gases~\cite{wenzel2017striped} have been observed separately, the coexistence of global coherence and crystalline structure - essential for supersolidity - has not been achieved in 2D systems. Theoretical quantum Monte Carlo simulations have predicted stripe phases in 2D~\cite{macia2012excitations,bombin2017dipolar}, but whether these stripes can be coherently coupled and exhibit supersolidity remains debated~\cite{bombin2019berezinskii,cinti2019absence}. Particularly interesting scenarios emerge when the dipoles are tilted, potentially leading to supersolid phase emergence with relatively low density and weak interaction strength~\cite{block2014properties,fedorov2014two,staudinger2023striped,aleksandrova2024density,sanchez2025tilted}. To date, however, the existence of supersolidity in 2D dipolar gases remains an open question~\cite{Recati2023SupersolidityIU}.


In this work, we address this long-standing question by experimentally realizing a supersolid stripe state in a 2D dipolar gas of erbium atoms. Our key innovation is the precise control of the dipole orientation, which allows us to induce an anisotropic roton instability that triggers crystallization while preserving the underlying superfluid coherence. We present evidences for supersolidity, including the stripe crystal with a fixed spatial period, the measurement of global phase coherence across the stripes, the characterization of phase rigidity, and superfluid hydrodynamics. The observed stripe state is long-lived, proving a new mechanism for the realization of a supersolid in genuine 2D systems. Moreover, our work opens the possibility of studying the 2D superfluid transition with exotic structures.

\begin{figure}
\begin{center}
\includegraphics[scale=0.44]{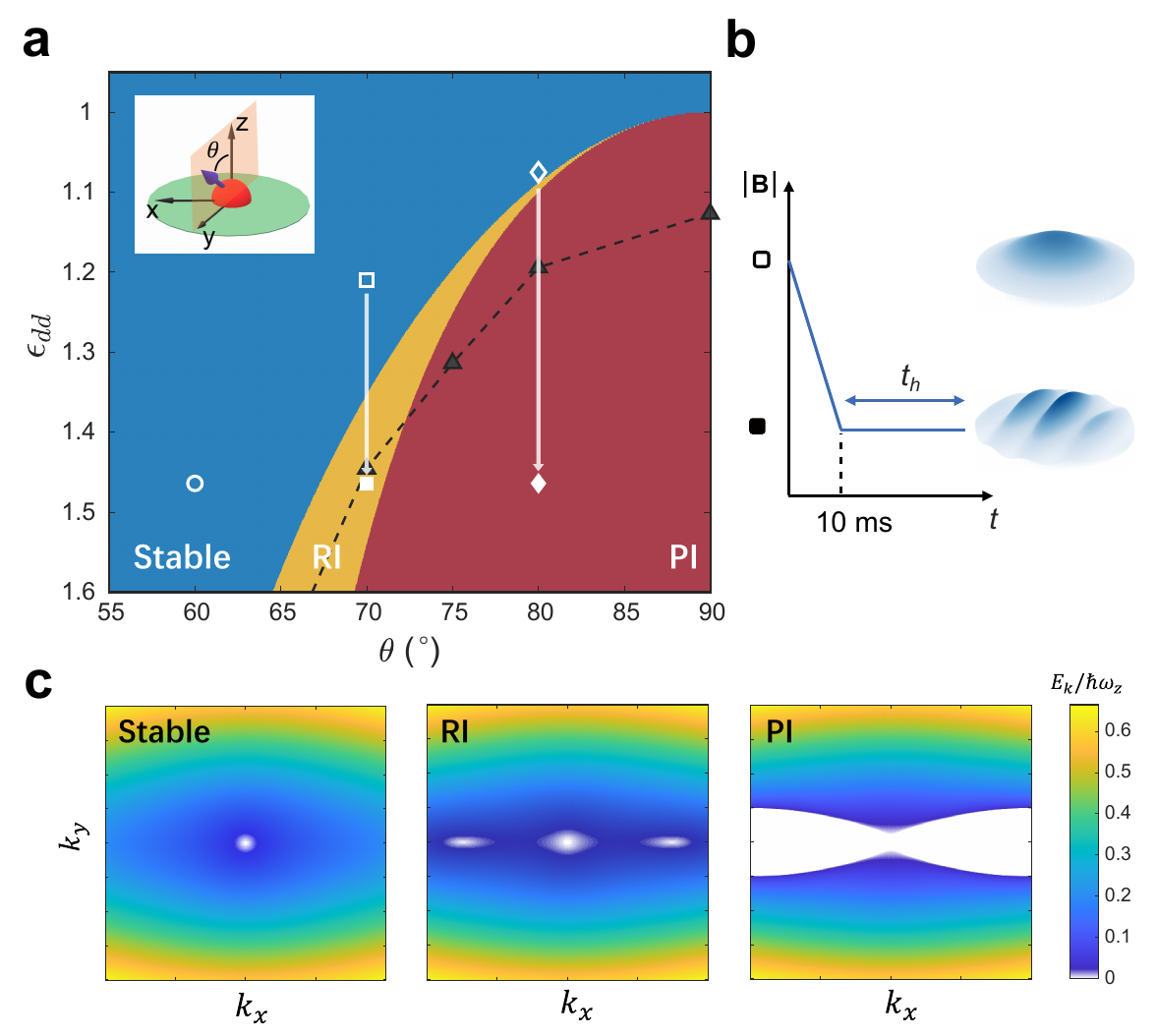}\\

\caption{\textbf{Stability diagram of 2D dipolar condensates.} \textbf{a.} Different colors denote different regimes calculated with a homogeneous 2D density $n_{2D}=100~\mu m^{-2}$ and $l_z=272$~nm by Eq.~\ref{Eq1} for erbium atoms. In the stable regime~(blue), $g_{\text{eff}}$ is well above zero and the excitation energy is always real. In the roton instability~(RI) regime~(yellow), an imaginary part emerges at finite momentum, while $g_{\text{eff}}$ is still positive. In the phonon instability~(PI) regime~(red), $g_{\text{eff}}<0$ so no stable phonon exists. Black triangles are the critical values for the formation of stripes, obtained by the teGPE simulations with our experimental sequences. A stripe state appears below the black triangles. \textbf{b.} Time sequence for sample preparation in unstable regimes. We first prepare stable superfluid samples with different dipolar angle $\theta$ close to the instability boundary denoted by open symbols in \textbf{a}, then linearly ramp the magnetic field in 10 ms to the final value denoted by solid symbols to enter unstable regimes, meanwhile keeping $\theta$ constant. \textbf{c.} Schematic 2D excitation spectrum in different regimes calculated from Eq.~\ref{Eq1}. Two roton minima are clearly visualized at $(\pm k_{\text{rot}},0)$ in the RI regime. The anisotropic roton triggers the crystallization along the $x$-axis and so the formation of stripes. 
}\label{fig1}
\end{center}
\end{figure}

\vspace{10pt}
{\textbf{Mean-field instabilities in 2D.}} Fig.~\ref{fig1}a shows different instability regimes for a quasi-2D dipolar condensate, which can be determined from Bogoliubov excitation as a function of dipole tilt angle $\theta$ and relative dipolar strength $\epsilon_{dd}=a_{dd}/a_s$ with s-wave scattering length $a_s$ and dipolar length $a_{dd}=\mu_0\mu_m^2m/12\pi\hbar^2$~\cite{mishra2016dipolar}. 
For a homogeneous 2D dipolar condensate with density $n$ and dipoles polarized at an angle $\theta$, the excitation energy is given by $E_k=\sqrt{\frac{\hbar^2k^2}{2m}\left[\frac{\hbar^2k^2}{2m}+2nV(\textbf{k})\right]}$, with the momentum-dependent interaction~\cite{ticknor2011anisotropic} 
\begin{equation}
    V(\textbf{k})=g_{\text{eff}}-3g_{d}G(kl_z/\sqrt{2})[\cos^2(\theta)-\cos^2{\phi_k}\sin^2(\theta)],
    \label{Eq1}
\end{equation}
$G(q)\equiv \sqrt{\pi}qe^{q^2} {\rm erfc} (q)$ with erfc the complementary error function and $\phi_k$ is the angle between $k$ and the projection of dipole orientation on the 2D plane. We introduce an effective local coupling constant $g_{\text{eff}}=g_s+g_{d}(3\cos^2\theta-1)$, $g_s=\sqrt{8\pi}\hbar^2 a_s/ml_z$ is the quasi-2D contact coupling constant and $g_{d}=\sqrt{8\pi}\hbar^2a_{dd}/ml_z$ is the quasi-2D dipolar coupling constant, with the axial harmonic oscillator length $l_z=\sqrt{\hbar/m\omega_z}$. The mean-field instability occurs when tuning $g_{\text{eff}}$ close to or below zero, where an imaginary part starts to emerge in the excitation spectrum. 


A schematic illustration of the real part of the 2D excitation spectrum in different regimes is shown in Fig.~\ref{fig1}c. One crucial mechanism is the roton instability (RI), which arises when the excitation energy becomes imaginary at a specific, non-zero momentum. For tilted dipoles, this anisotropic instability creates roton minima in the momentum space, forming striped density patterns in a single direction without requiring anisotropic confinement. Since a mean-field theory alone predicts these stripes to be transient~\cite{mishra2016dipolar}, the inclusion of beyond-mean-field interactions appears to be critical to form a long-lived supersolid stripe crystal~\cite{aleksandrova2024density,sanchez2025tilted}. We further perform time-dependent extended Gross-Pitaevskii equation~
(teGPE) simulations which include the beyond mean-field Lee-Huang-Yang~(LHY) correction and the external harmonic trap, and suggests the metastability of the stripe state (see supplementary material). The critical points of stripe formation in the real-time simulation of the experimental protocol described below are given by the black triangles in Fig.~\ref{fig1}a~\cite{SI}. The repulsive LHY correction pushes the instability boundary to slightly larger $\epsilon_{dd}$ compared to the mean-field picture.

\vspace{10pt}
{\textbf{Experiments.}}
Our experiments are conducted with the $^{166}$Er isotope, with a large magnetic moment $\mu_m=7\mu_B$, where $\mu_B$ is the Bohr magneton. We typically prepare $32,000$ atoms at around $30$~nK in a quasi-2D trap with trap frequencies $(\omega_x, \omega_y, \omega_z)=2\pi\times(14.3(5),15.9(5),820(7))$ Hz at equilibrium in the stable regime with different $\theta$ as denoted by open symbols in Fig.~\ref{fig1}a~\cite{he2025exploring,seo2023apparatus}. The tilt angle $\theta$ with respect to the $z$ axis is set by the orientation of the external magnetic field in the $y-z$ plane. The dipolar length $a_{dd}=65.5 a_0$ is fixed for erbium atoms while $a_0$ is the Bohr radius. $\epsilon_{dd}$ can be adjusted by changing $a_s$ through the magnitude of the magnetic field close to the Feshbach-resonance at 0~G~\cite{patscheider2022determination}. The condensation fraction of the initial stable samples is 60\%, evaluated by a bimodal fitting of the TOF image from the side. The vertical confinement $\hbar\omega_z$ corresponds to $k_B\times 40$~nK while the chemical potentials of all initial samples are estimated to be lower than $k_B\times 10$~nK from the eGPE, so the quasi-2D condition is fulfilled. The quasi-2D nature is further confirmed by the 3D teGPE simulations, where we observe that the dynamics along $z$ are frozen to the ground state wave-function~\cite{SI}. The temperature is well below the critical temperature for a non-interacting ideal BEC in a 2D harmonic trap $T_c=\left(\frac{6N}{\pi^2}\right)^{\frac{1}{2}}\frac{\hbar\omega_r}{k_B}\sim100$~nK, where $\omega_r=\sqrt{\omega_x\omega_y}$ is the averaged planar trap frequency, so the initial samples are in the superfluid state~\cite{he2025exploring}.


\begin{figure*}
\begin{center}
\includegraphics[scale=0.52]{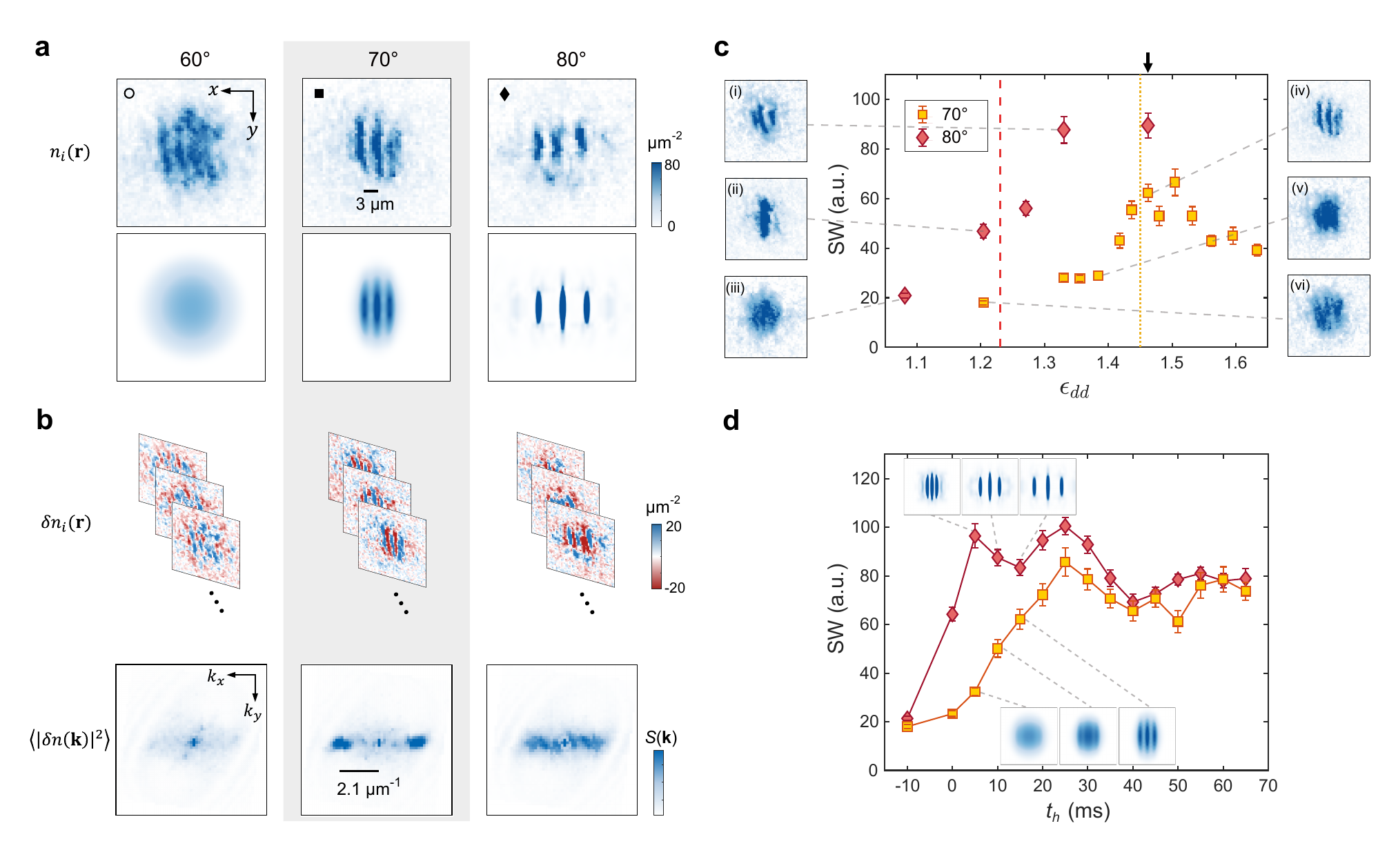}\\

\caption{\textbf{\textit{In-situ} measurements of crystallization of a 2D dipolar superfluid.} \textbf{a.} Upper row: exemplary \textit{In-situ} images of samples at 180 mG~($\epsilon_{dd}=1.46$), $t_h=15$~ms with $\theta=60^\circ,70^\circ,80^\circ$ respectively. The interaction parameters correspond to the bottom symbols in Fig.~\ref{fig1}a. Bottom row: 2D density profiles given by teGPE simulation. The saturating density in the color scale is adjusted to be better compared with the experimental data. \textbf{b.} Static structure factor measured by density fluctuations. \textbf{c.} SW as a function of final $\epsilon_{dd}$, showing that 2D dipolar superfluid start to crystallize after entering unstable regime. The vertical dotted~(dashed) line denote the critical value in Fig.~\ref{fig1}a for $\theta=70^\circ$~($80^\circ$). The black arrow denotes $\epsilon_{dd}=1.46$ at 180 mG where we focus on. (i)-(iii), exemplary single images of samples at $\theta=80^\circ$. (iv)-(vi), exemplary single images of samples at $\theta=70^\circ$. \textbf{d.} Formation dynamics of crystalline order at $\epsilon_{dd}=1.46$, $\theta=70^\circ,80^\circ$. Insets are simulated 2D profiles at $t_h=5,10,15$ ms. Errorbars correspond to the standard error of the mean.}\label{fig2}
\end{center}
\end{figure*}

Starting from a stable sample, we linearly ramp down the magnetic field in order to decrease $a_s$ in 10~ms to the target value~(typically 180~mG, corresponding to $\epsilon_{dd}=1.46$ as denoted by solid symbols in Fig.~\ref{fig1}a) to enter the unstable regime while keeping $\theta$ constant. Then we hold for $t_h$~(typically 15~ms if not specified) to ensure the formation of density modulations, as shown in Fig.~\ref{fig1}b. The density distributions of atoms are imaged after by high-intensity absorption imaging along the $z$ direction with a spatial resolution of approximately 1~$\mu$m.

\vspace{10pt}
{\textbf{Crystallization of a 2D dipolar superfluid.}}
We present the exemplary \textit{in-situ} density distribution at $t_h=15$ ms with dipoles tilted to $\theta=60^\circ, 70^\circ, 80^\circ$ at $\epsilon_{dd}=1.46$, alongside profiles generated by the teGPE simulation in Fig.~\ref{fig2}a. High-contrast stripes are clearly visible in both the PI ($\theta=80^\circ$) and RI ($\theta=70^\circ$) regimes. In the simulation, the stripe crystals at $70^\circ$ are connected through a pronounced density in between, giving a large superfluid fraction according to Leggett's bounds~\cite{leggett70:prl,leggett98:jsph,sanchez2025tilted} ($f_s = 0.90 \pm 0.01$). On the other hand, the stripes at $\theta = 80^\circ$ are disconnected in the simulation, implying that global phase coherence is absent. Near the instability boundary at $\theta=60^\circ$, subtle vertical-stripe structures appear intermittently between experimental shots~\cite{zhen2025breaking}. These fluctuating structures differ significantly from the clear stripes observed in unstable regimes. The vertical structure at $60^\circ$ represents anisotropic density fluctuation~\cite{he2025exploring,baillie2014number}—a precursor to fully formed stripe states.

\begin{figure*}
\begin{center}
\includegraphics[scale=0.58]{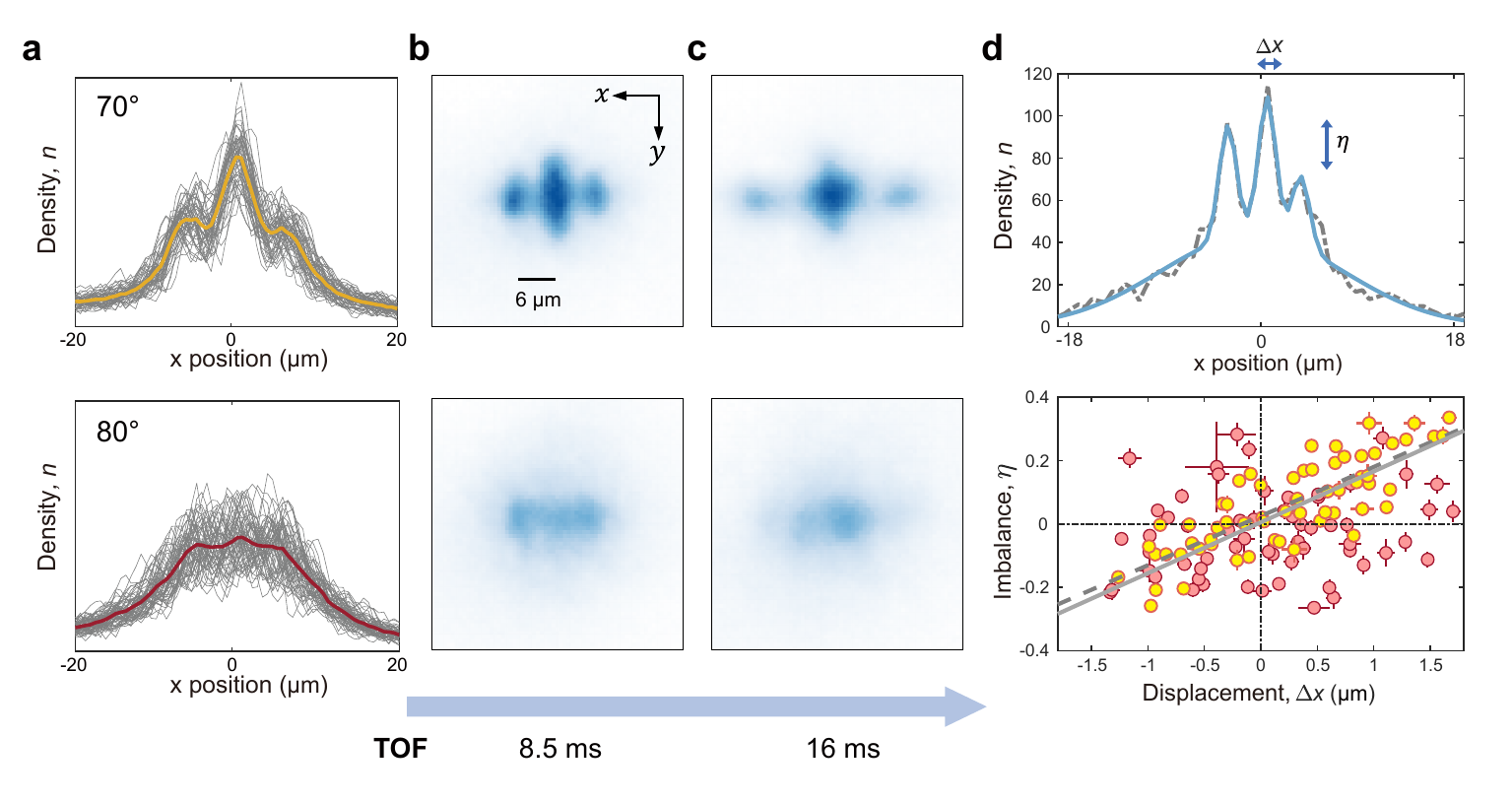}\\

\caption{\textbf{Probing phase coherence and phase rigidity of 2D dipolar stripes.} \textbf{a.} Integrated density distribution after 8.5 ms TOF interference, gray lines are individual measurements, yellow and red lines correspond to the average. Clear density modulation of TOF images is preserved at $\theta=70^\circ$ after averaging more than 50 trials, showing the global coherence. Meanwhile, interference patterns with random distribution are observed at $\theta=80^\circ$. \textbf{b,c} Averaged 2D density distribution after 8.5 ms and 16 ms TOF~(averaged more than 50 trials). Upper panel for $\theta=70^\circ$ and lower panel for $\theta=80^\circ$. \textbf{d.} Upper panel: the gray dashed line corresponds to the 1D \textit{in situ} density distribution integrated along the $y$ direction with three stripes. The blue solid line is the fitting profile to extract the displacement $\Delta x$ and the imbalance $\eta$. Lower panel: $\Delta x$ \textit{vs.} $\eta$ at $70^\circ$~(yellow) and $80^\circ$~(red). Samples at $70^\circ$ show linear $\Delta x-\eta$ correlations, signaling an effective supersolid low-energy Goldstone mode. The gray dashed line is the linear fit of the experimental data at $\theta=70^\circ$. The gray solid line is the $\Delta x-\eta$ linear correlation obtained from teGPE simulations~\cite{SI}. Errorbars represent fitting errors.}\label{fig3}
\end{center}
\end{figure*}


The emergent lattice structure can be revealed by analyzing the density fluctuation $\delta n_i(\textbf{r})=n_i(\textbf{r})-\langle n(\textbf{r})\rangle$. We obtain the power spectrum of this density fluctuation through a 2D fast Fourier transform, $\delta n_i(\textbf{k})=FFT[\delta n_i(\textbf{r})]$, resulting in the static structure factor $S(\textbf{k})=\langle|\delta n(\textbf{k})|^2\rangle/N$ as shown in Fig.~\ref{fig2}b, where N represents the mean atom number~\cite{hertkorn2021density}. Each $S(\textbf{k})$ is obtained by averaging over 50 individual measurements. At $\theta=60^\circ$, the static structure factor is a dim ellipse along $x$, signaling the anisotropic static structure factor in a 2D dipolar superfluid with tilted dipoles~\cite{macia2012excitations,baillie2014number}. At $\theta=70^\circ$, which is in the RI regime, two peaks at $(\pm k_{rot}, 0)$ are observed in $S(\textbf{k})$. The two peaks signal the enhanced density-density correlations at the roton minimum~\cite{macia2012excitations,hertkorn2021density}. It also indicates that the emergent lattice has a stable spatial period along the $x$ direction. At $\theta=80^\circ$, stripes with random intervals are observed, resulting in a blurred band in $S(\textbf{k})$. This can be understood by the fact that there is no characteristic length scale along the $x$ axis at $80^\circ$ in the excitation spectrum in Fig.~\ref{fig1}c.

To quantify the crystalline order along the $x$ axis, we define the weight of static structure factor $\text{SW}=\sum\limits_{k_x=1~\mu \text{m}^{-1}}^{3~\mu \text{m}^{-1}} \sum\limits_{k_y}S(\textbf{k})$ which is a quantity describing the strength of the striped density modulations. We measure SW at $t_h=15$ ms with increasing $\epsilon_{dd}$ at $70^\circ$ and $80^\circ$ as shown in Fig.~\ref{fig2}c. When entering the PI regime with $g_{\text{eff}}<0$ at $80^\circ$, the cloud first becomes an elongated single stripe, indicating a macrodroplet state in 2D, with multiple stripes arising if $\epsilon_{dd}$ is furtherly increased. When entering the PI regime with $g_{\text{eff}}>0$ at $70^\circ$, stripes with a constant spatial period emerge on top of a superfluid background. The critical $\epsilon_{dd}$ for the formation of stripes show good agreement with the teGPE simulations as denoted by the vertical lines in Fig.~\ref{fig2}c. We focus on the final value $\epsilon_{dd}=1.46$ at $B=180$ mG and monitor the formation dynamic of the striped crystal as shown in Fig.~\ref{fig2}d. The formation of stripes at $80^\circ$ turns out to be faster than at $70^\circ$, which is correctly captured by the teGPE simulation at the early stage as shown in the insets of Fig.~\ref{fig2}d. The density modulations persist for several tens of milliseconds.

\vspace{10pt}
{\textbf{Supersolidity of 2D dipolar stripes.}} To investigate the coherence properties of the 2D stripe states, we perform standard TOF matter-wave interference measurements by suddenly turning off the trap to let the stripes freely expand and interfere. We rotate the magnetic field to the $z$ direction at $1$ G within the first $0.5$~ms during TOF to accelerate the expansion of the stripes. The periodic interference patterns are observed in each individual measurement at both $70^\circ$ and $80^\circ$, showing that each stripe is a coherent sample. At $70^\circ$, the TOF profile shows a clear modulation after averaging over 50 images, as shown in Fig.~\ref{fig3}a upper panel. We confirm that the interval between side peaks are scaled linearly with the TOF time as shown in Fig.~\ref{fig3}b,c upper panel. This observation provides the evidence of global phase coherence of the stripes in the explored RI regime, which makes it a supersolid stripe. In contrast, the interference pattern shows large shot-to-shot fluctuations at $\theta=80^\circ$ and no modulation is observed in the average image as shown in Fig.~\ref{fig3}a lower panel, indicating that there is no global coherence.


A supersolid state relies on particle exchange through a superfluid flow between the stripes to maintain global coherence, which gives rise to a low-energy 
Goldstone mode~\cite{guo2019low}. This means that the phase shift of the crystal structure is compensated by superfluid flow. Typically, we observe $2\sim3$ stripes at $\theta=70^\circ$ and $3\sim4$ stripes at $\theta=80^\circ$, with $B=180$ mG. We post-select data with 3 stripes to analyze the correlation between displacement from center of mass $\Delta x$ and the imbalance $\eta$ in the stripe states. We fit the integrated density profile along $y$ with a sum of four Gaussian functions~(for three stripes and the background) as shown in the upper panel of Fig.~\ref{fig3}d. From the fitting, we extract the position and atom number of the $i_{th}$ stripe $x_i$ and $N_i$~($i$ from left to right), then determine displacement $\Delta x=(x_1+x_2+x_3)/3-x_{COM}$ and imbalance $\eta=(N_1-N_3)/(N_1+N_2+N_3)$~\cite{guo2019low}. The cumulative measurement result is shown by the lower panel of Fig.~\ref{fig3}d. The stripes in the RI regime show a significant linear $\Delta x-\eta$ correlation, which establishes the phase rigidity of the supersolid samples at $\theta=70^\circ$. The correlation observed in the experiment closely agrees with the teGPE simulation. Again, the linear correlation of $\Delta x$ and $\eta$ diminishes in the PI regime at $\theta=80^\circ$.

\begin{figure}
\begin{center}
\includegraphics[scale=0.8]{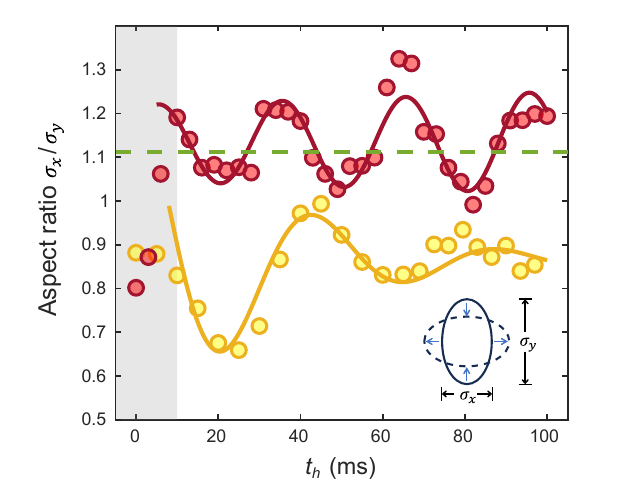}\\

\caption{\textbf{Superfluid hydrodynamics of a 2D dipolar supersolid.} Yellow (red) circles correspond to the aspect ratio of averaged \textit{in situ} images at different $t_h$ with $\theta=70^\circ (80^\circ)$, fitting errors are smaller than the symbol size. The periodic oscillations of the AR indicate a collective quadrupole mode excited in the system. Solid curves are damped sinusoidal fittings starting from $t_h=10$~ms. The green dashed line denotes the aspect ratio of the trap $\omega_y/\omega_x$.}\label{fig4}
\end{center}
\end{figure}

\vspace{10pt}
{\textbf{Hydrodynamics collective behavior.}} 
The superfluid nature of the $70^\circ$ sample is further revealed by a collective oscillation of the cloud induced when we ramp down the magnetic field to decrease $a_s$. We identify the predominantly excited mode as the quadrupole mode, resulting from the anisotropic DDI. After ramping the magnetic field to the target value, we monitor the $\textit{in-situ}$ size of the major cloud at different hold times $t_h$ by averaging many images, effectively washing out the density modulation structures in the center. Using a 2D Gaussian fitting, we extract the widths along the $x$-axis ($\sigma_x$) and $y$-axis ($\sigma_y$). We then plot the evolution of the aspect ratio (AR) $\sigma_x/\sigma_y$ in Fig.~\ref{fig4}.

Strikingly different collective behavior is observed between $70^\circ$(RI) and $80^\circ$(PI) samples at $\epsilon_{dd}=1.46$. In the RI regime, the aspect ratio $\sigma_x/\sigma_y$ oscillates at a frequency $\omega_Q/\omega_{r}=1.5(1)$, which is close to the hydrodynamic limit $\sqrt{2}$~\cite{de2015collective}. This indicates that the striped crystal created in the RI regime is a single coherent sample exhibiting superfluid hydrodynamic behavior. In the PI regime, however, the aspect ratio oscillates at a frequency $\omega_Q/\omega_{r}=2.1(1)$, which approaches the collisionless limit of $2$. This signals that when the striped crystal forms in the PI regime, the stripes remain uncoupled without maintaining coherence. This distinct collective oscillation behavior at $\theta=70^\circ$ and $\theta=80^\circ$ is also observed in our teGPE simulation~\cite{SI}.

The absolute values of ARs also give a hint of the coherence of the background cloud. Although our planar trap has an aspect ratio $\omega_y/\omega_x=1.11$, which is slightly elongated along the $x$-axis, the shape of the cloud is elongated along the $y$-axis~($\sigma_x/\sigma_y<1$) in the stable superfluid regime due to the magnetostriction effect, which is expected to be only significant in the superfluid state. The magnetostriction is preserved for 100 ms at $70^\circ$, since the AR of the cloud is always smaller than $1$ during the oscillation, as shown in Fig.~\ref{fig4}. Meanwhile, at $80^\circ$, starting from $AR<1$ at $t_h=0$~ms, the AR rapidly increases to larger than $1$ within $10$~ms. Later, it oscillates around $1.1$ which is the AR of the trap $\omega_y/\omega_x$. The inverting of the AR evidences that the background cloud rapidly loses global coherence when entering into the deep PI regime.

Both the frequency of the quadrupole mode and the aspect ratio of the background cloud suggest that when $\epsilon_{dd}=1.46$, at $70^\circ$(RI) the stripes are embedded in a 2D superfluid background , while at $80^\circ$(PI) the stripes are embedded in an incoherent gas.

\vspace{6pt}
{\textbf{Conclusion.}} In this work, we have observed a long-lived stripe state emerging in 2D dipolar gases. Our comprehensive measurements demonstrate the hallmarks of this state including crystalline order with stable lattice period, global coherence, low-energy Goldstone mode, and hydrodynamic collective mode , all of which reveal the supersolid nature in our 2D system. We find that the tilting angle $\theta$ of the dipoles serves as a crucial and extremely sensitive parameter for realizing the supersolid state, opening new opportunities for exploring novel phases of matter. In dilute 2D dipolar gases, maintaining the system in the RI regime (where the local coupling $g_{\text{eff}}$ remains positive) provides a broader window for crystalline structures to preserve superfluidity. Since $g_{\text{eff}}$ determines the BKT transition point in stable weakly-interacting 2D dipolar gases away from unstable regimes~\cite{he2025exploring}, it likely plays a crucial role in maintaining 2D superfluidity in the low-density regions connecting the stripes.

Our work opens the door to the further exploration of supersolidity in two dimensions in a controllable environment. In particular, a promising direction for future research involves the experimental investigation of the interplay between the BKT transition~\cite{bombin2019berezinskii}, expected to emerge in a 2D setting, and supersolidity. Since a system of dipoles in a quasi-2D geometry features continuous, as well as discontinuous transitions between the homogeneous and the stripe phases~\cite{aleksandrova2024density,sanchez2025tilted}, our setup paves the way for exploring the Kibble-Zurek mechanism on the dynamics of the supersolid transition via the quenching of the interaction. Moreover, since our set-up lies in the strongly dipolar regime, it is an ideal candidate to study the proliferation and dynamics of vortices and their unconventional pairing~\cite{mulkerin2013anisotropic} in a supersolid constrained to lower dimensionality.


\vspace{10pt}
{\bf Acknowledgement}
GBJ acknowledges support from the RGC through RFS2122-6S04. J.S-B and J.B.  acknowledge support by the Spanish Ministerio de Ciencia, Innovación y Universidades (grant PID2023-147469NB-C21, financed by MICIU/AEI/10.13039/501100011033 and FEDER-EU). N.D acknowledges support by the Swiss National Science Foundation (SNSF) grant numbers 200021 207537 and 200021 236722 and the Swiss State Secretariat for Education, Research and Innovation (SERI).

\newpage
\providecommand{\noopsort}[1]{}\providecommand{\singleletter}[1]{#1}%

\newpage

\end{document}


\renewcommand*{\thefigure}{S\arabic{figure}}

\preprint{APS/123-QED}

\title{Supplementary note : Observation of a supersolid stripe state in two-dimensional
dipolar gases}

\author{Yifei He}
\affiliation{Department of Physics, The Hong Kong University of Science and Technology,
Clear Water Bay, Kowloon, Hong Kong}%

\author{Haoting Zhen}
\affiliation{Department of Physics, The Hong Kong University of Science and Technology,
Clear Water Bay, Kowloon, Hong Kong}%

\author{Mithilesh K. Parit}
\affiliation{Department of Physics, The Hong Kong University of Science and Technology,
Clear Water Bay, Kowloon, Hong Kong}%

\author{Mingchen Huang}
\affiliation{Department of Physics, The Hong Kong University of Science and Technology,
Clear Water Bay, Kowloon, Hong Kong}%

\author{Nicol\`o Defenu}
\affiliation{Institut f\"ur Theoretische Physik, ETH Z\"urich, Wolfgang-Pauli-Str. 27 Z\"urich, Switzerland.}%

\author{Jordi Boronat}
\affiliation{Departament de Física, Universitat Politècnica de Catalunya, Campus Nord B4-B5, 08034 Barcelona, Spain}%

\author{Juan Sánchez-Baena}
\affiliation{Departament de Física, Universitat Politècnica de Catalunya, Campus Nord B4-B5, 08034 Barcelona, Spain}%

\author{Gyu-Boong Jo}
\altaffiliation{email: gbjo@rice.edu}
\affiliation{Department of Physics and Astronomy, Rice University, Houston, TX, USA}%
\affiliation{Smalley-Curl Institute, Rice University, Houston, TX, USA}%
\affiliation{Department of Physics, The Hong Kong University 
of Science and Technology,
Clear Water Bay, Kowloon, Hong Kong}%

\date{\today}

\maketitle


\newpage

\subsection{Sample preparation and experimental detail}

Our experiments start from almost pure Bose-Einstein-condensates~(BEC) of $^{166}$Er in a crossed optical dipole trap~(cODT) formed by two laser beams at 1064 nm laser. Then we transfer the atoms from the cODT to a quasi-2D trap created by a light sheet and a vertical ODT with 532~nm laser, with trap frequency $(f_x, f_y, f_z)=(14.3(5),15.9(5),820(7))$ Hz. We adiabatically change the direction and magnitude of the magnetic field to the target value, creating stable superfluid samples in equilibrium following the procedure in our previous work~\cite{he2025exploring}. The initial samples are prepared under magnetic field $B_i=~$180 mG, 300 mG, 500 mG for $\theta=60^\circ,70^\circ,80^\circ$ respectively, correspond to $\epsilon_{dd}=1.46, 1.20, 1.08$. We use traditional bimodal fitting on the images after 16 ms TOF from side to evaluate the temperature and condensation fraction of our initial samples.

Starting from the initial samples in the stable regime, we ramp the magnetic field at $70^\circ$ and $80^\circ$ to enter different instability regimes following the sequences described in the main text, utilizing a Feshbach resonance close to 0 G~\cite{patscheider2022determination}. After we prepare the samples in target states, we use a high-power absorption image that is performed along the $z$ axis using an objective with a numerical aperture of 0.28, with resolution approximately $1~\mu$m, to probe the density distribution of atoms. The power of the image beam $I/I_{sat}=20$ and the duration of the image pulse is $5~\mu$s.

The magnitude and orientation of the external magnetic field are controlled by three orthogonal pairs of Helmholtz coils, whose currents are stabilized by PID circuits. We produce all the erbium atoms in their lowest Zeeman level $m_J=-6$, so the orientation of the dipoles adiabatically follows the orientation of the external magnetic field. The contact coupling strength $g_s$ is controlled by s-wave scattering length $a_s$ in our experiment. The magnetic field-dependent $a_s$ is modeled as $a_s=(a_{bg}+sB)\prod \limits_i(1-\frac{\Delta B_i}{B-B_i})$
where B is the magnetic field strength, $i$ denotes $i^{th}$ Feshbach resonances. Using the latest calibration result of $a_{bg}$, $s$, $B_i$ and $\Delta B_i$ for $^{168}$Er~\cite{patscheider2022determination}, we estimate $a_s$ at given magnetic field strengths. The uncertainty of the magnetic field strength is $\pm2$~mG calibrated by radio frequency spectroscopy at 180 mG, corresponding to a $\pm0.5a_0$ uncertainty in $a_s$ and $\pm0.4^\circ$ in $\theta$. Other than that, the calibration uncertainty of the background scattering length contributes to a systematic uncertainty $\pm3a_0$ in $a_s$~\cite{patscheider2022determination}.

\subsection{Arbitrary stripe orientation induced by tilted dipoles}
\begin{figure}
\centering
\includegraphics[width=0.85\linewidth]{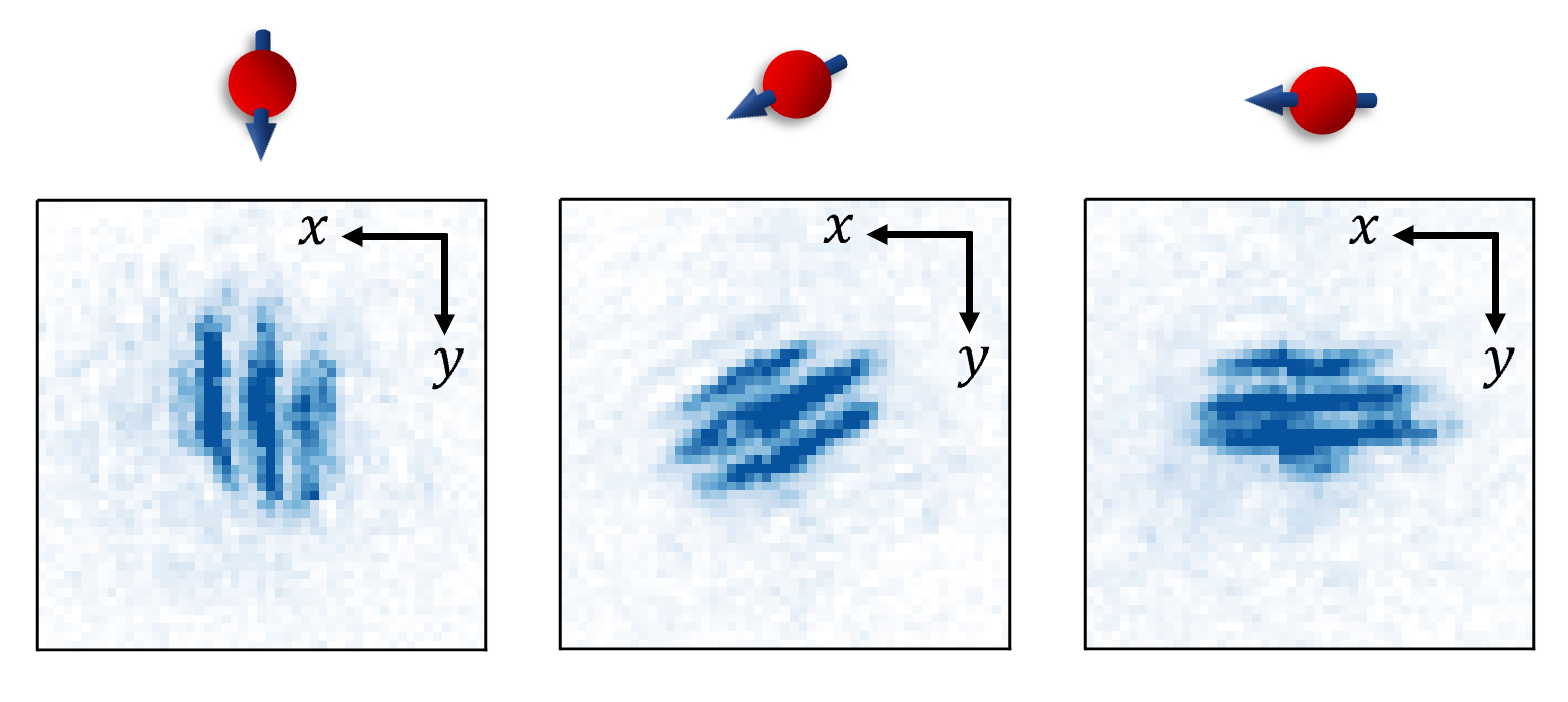}
\caption{ Exemplary \textit{in situ} images of supersolid stripes along different directions, controlled by the azimuthal angle of the dipoles.
}
\label{figS1}
\end{figure}

Despite the slight anisotropy of the external harmonic trap, the formation of the striped density wave is mostly dependent on the orientation of dipoles, which intrinsically breaks the rotational symmetry. We show that by changing the azimuthal angle of the dipoles, we can produce stripe crystals with a similar lattice period along any direction in the RI regime, following the same sequence described in the main text, as shown in Fig.~\ref{figS1}. This observation signals that in the 2D plane an anisotropic roton spectrum is induced by tilting dipoles, where the roton minimum is determined by the azimuthal angle of the dipoles. This feature further establishes the 2D nature of the excitations in our system.

\subsection{teGPE simulations}

In order to simulate the experimental protocol to generate the quantum states shown in Fig.~2 of the main text, we solve numerically the time-dependent extended Gross-Pitaevskii equation (teGPE), which is given by
\begin{align}
 i \hbar \pdv{\psi}{t} =& \left[-\frac{\hbar^2\nabla^2}{2m} + U({\bf r}) + H_{\rm qu}({\bf r},t)
 \right. \nonumber \\
 &\left. + \!\!\int \!{\rm d}{\bf r}^\prime V({\bf r}-{\bf r}') \left( \abs{\psi({\bf
r}',t)}^2 \right) \right] \psi({\bf
r},t) \label{teGPE} \\
 &= \hat{K} \psi({\bf r},t) + \hat{V}_{\rm teGPE}({\bf r},t) \psi({\bf r},t) \ .
\end{align}
In this expression, $\psi({\bf r},t)$ is the condensate wave function, $m$ is the atomic mass, the term $U({\bf r}) = \frac{1}{2} m \left( \omega_x^2 x^2 + \omega_y^2 y^2 + \omega_z^2 z^2 \right)$ is the harmonic trap with frequencies $(\omega_x,\omega_y,\omega_z)=2\pi \times (14.3, 15.9, 820)$ Hz while $\hat{K} = \frac{\hat{P}^2}{2 m}$ is the kinetic energy operator. The interaction
$V({\bf r})$ is given by
\begin{align}
 V({\bf r} - {\bf r'}) = \frac{4 \pi \hbar^2 a_s}{m} \delta\left( {\bf r} - {\bf r'} \right) + \frac{C_{\rm dd}}{4 \pi} \frac{ \left( 1 - 3 \cos^2 \theta_{\alpha} \right) }{\abs{{\bf r} - {\bf r'}}^3} \ , \label{pseudopot}
\end{align}
with $a_s$ the s-wave scattering length, $C_{\rm dd} = \mu_0 \mu_m^2$ for magnetic dipoles and $\theta_{\alpha}$ the angle between the vector ${\bf r} - {\bf r'}$ and the polarization axis of the dipoles. In all the cases considered in this work, $a \ll l_z$ with $l_z$ the harmonic oscillator length along the $z$-axis, which validates the model interaction of Eq.~\ref{pseudopot}. $H_{\rm qu}({\bf r},t)$ is the usual beyond mean-field correction employed in three-dimensions, which is given by
\begin{equation}
 H_{\rm qu}({\bf r},t) = \frac{32}{3 \sqrt{\pi}} g \sqrt{a^3} Q_5(a_{\rm dd}/a)
\abs{ \psi({\bf r},t) }^3 \ .
\end{equation}
The parameter $a_{\rm dd} = m \mu_0 \mu_m^2/(12\pi \hbar^2)$ corresponds to the dipole length ($a_{\rm dd} = 65.5 a_0$ for $^{166}$Er atoms), and the
auxiliary function $Q_5(a_{\rm dd}/a)$ is given by~\cite{pelster:PRA_2012}
\begin{equation}
 Q_5(a_{\rm dd}/a_s) = \int_{0}^1 du \left( 1 - \frac{a_{\rm dd}}{a_s} + 3
\left( \frac{a_{\rm dd}}{a_s} \right) u^2 \right)^{5/2} \ .
\end{equation}
The number of particles is set to $N=32000$ in the simulations.

Despite the high trapping strength applied along the $z$-axis and the quasi-2D 
character of the system, the densities of our system lay in the regime where the 
LHY correction accounting for the discretization of the excitations along the 
$z$ axis, due to the trapping potential, is close  
to its value in the fully 3D case~\cite{sanchez2025tilted}, which justifies the 
use of the standard 3D expression for the quantum fluctuations.

The protocol to obtain the results of Figs.~1a,~2 and~3 in the main text corresponds to a 10ms linear ramp of the scattering length from an initial value to a final one plus a holding time of 15ms. The initial and final values of the parameter $\epsilon_{\rm dd}$ in the simulations are given by $\epsilon^{\rm initial}_{\rm dd} = 1.2$, $\epsilon^{\rm final}_{\rm dd} = 1.466$ for $\theta = 70^o$ and by $\epsilon^{\rm initial}_{\rm dd} = 1.03$, $\epsilon^{\rm final}_{\rm dd} = 1.45$ for $\theta=80^o$, consistent with the experimental values within the experimental uncertainty in the scattering length. The initial state of the real time simulations is obtained by solving the imaginary time version of Eq.~\ref{teGPE} for $\epsilon_{\rm dd} = \epsilon^{\rm initial}_{\rm dd}$. In order to implement the real-time evolution, we apply the time-evolution operator to the wave function iteratively, which we split as
\begin{align}
 \hat{O}_t \simeq \exp\left[ -i \frac{ \hat{V}_{\rm teGPE} \Delta t }{ 2 \hbar} \right] \exp\left[ -i \frac{ \hat{K} \Delta t }{ \hbar} \right] \exp\left[ -i \frac{ \hat{V}_{\rm teGPE} \Delta t }{ 2 \hbar} \right] \ ,
\end{align}
which is exact up to order $(\Delta t)^2$. If the time step is small enough, we 
can use the initial state at every iteration to compute $V_{\rm teGPE}({\bf 
r},t)$, which depends on the density. The terms of the propagator that contain 
the potential are applied in position space, while the kinetic term is applied 
in momentum space by using a fast Fourier transform algorithm.

\begin{figure}[t]
\centering
\includegraphics[width=0.55\linewidth]{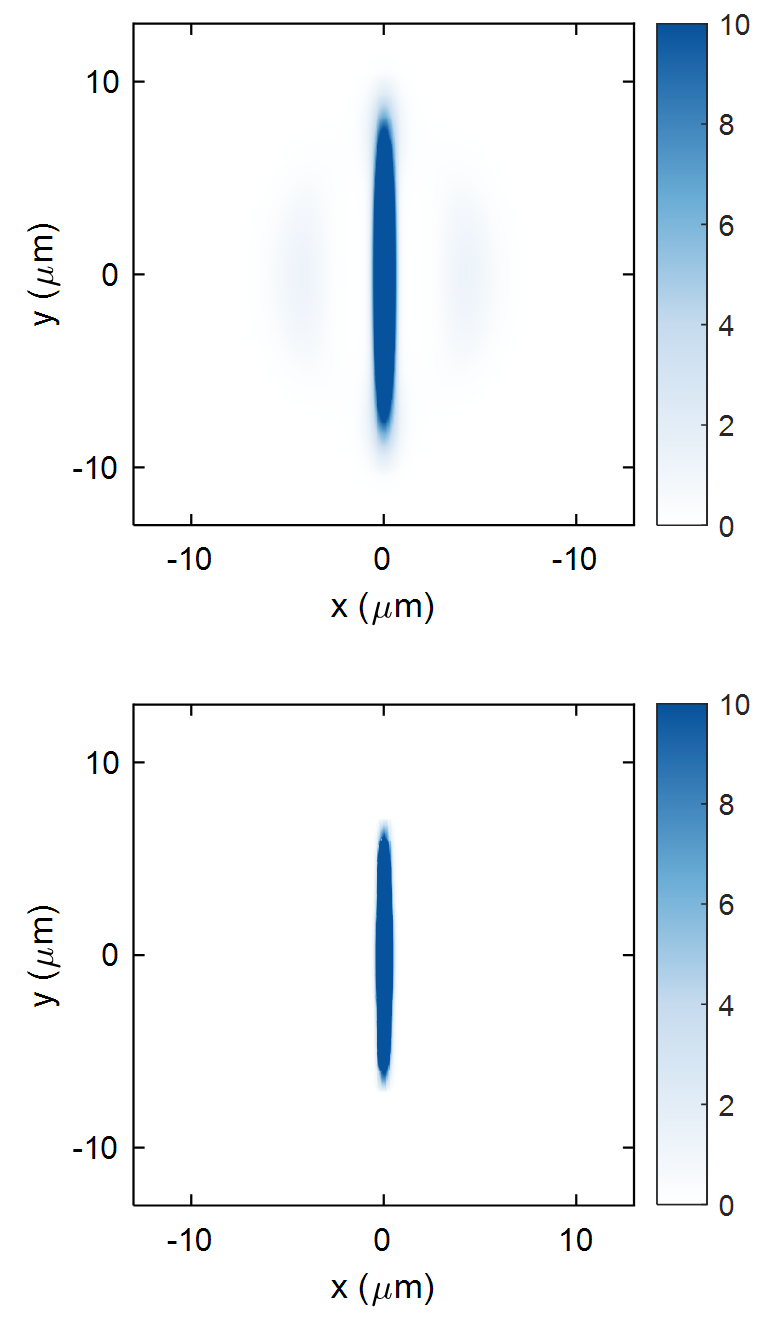}
\caption{ Ground state column densities $\rho (x,y) = \int dz \abs{\psi({\bf r})}^2$ corresponding to the ground state for $\theta= 70^o$ (top) and $\theta=80^o$ (bottom) at $\epsilon_{dd}=1.46$.
}
\label{figS2}
\end{figure}

\subsection{Ground state of the stripes phase}

Eq.~\ref{teGPE} can be propagated in imaginary time, which gives access to the ground state of the system for a given set of parameters. Interestingly, solving the eGPE in imaginary time reveals that the states shown in Fig.~2 of the main text are metastable states, while the ground state of the system for this parameters is given by an elongated gas for $\theta=70^o$ and an elongated, self-bound droplet for $\theta=80^o$, as shown in Fig.~\ref{figS2}. This is furtherly confirmed by looking at the energies of each configuration: the states produced in the experiment have an energy per particle of $E/N = 0.138 \epsilon$ for $\theta= 70^o$ and $E/N = 0.116 \epsilon$ for $\theta= 80^o$, while the solutions of the eGPE in imaginary time shown in Fig.~2 have an energy per particle of $E/N = 0.133 \epsilon$ and $E/N = -0.043 \epsilon$, for $\theta = 70^o$, $80^o$ respectively. Here, $\epsilon = \hbar^2/(m (12 \pi a_{\rm dd})^2)$ is a characteristic unit of energy. Interestingly, the supersolid state we produce following the experimental sequence has an energy per particle very close to the ground state.

\subsection{\label{sec:app-correlations}Extracting the $\eta-\Delta x$ correlations through the teGPE}

\begin{figure}[t]
\centering
\includegraphics[width=0.65\linewidth]{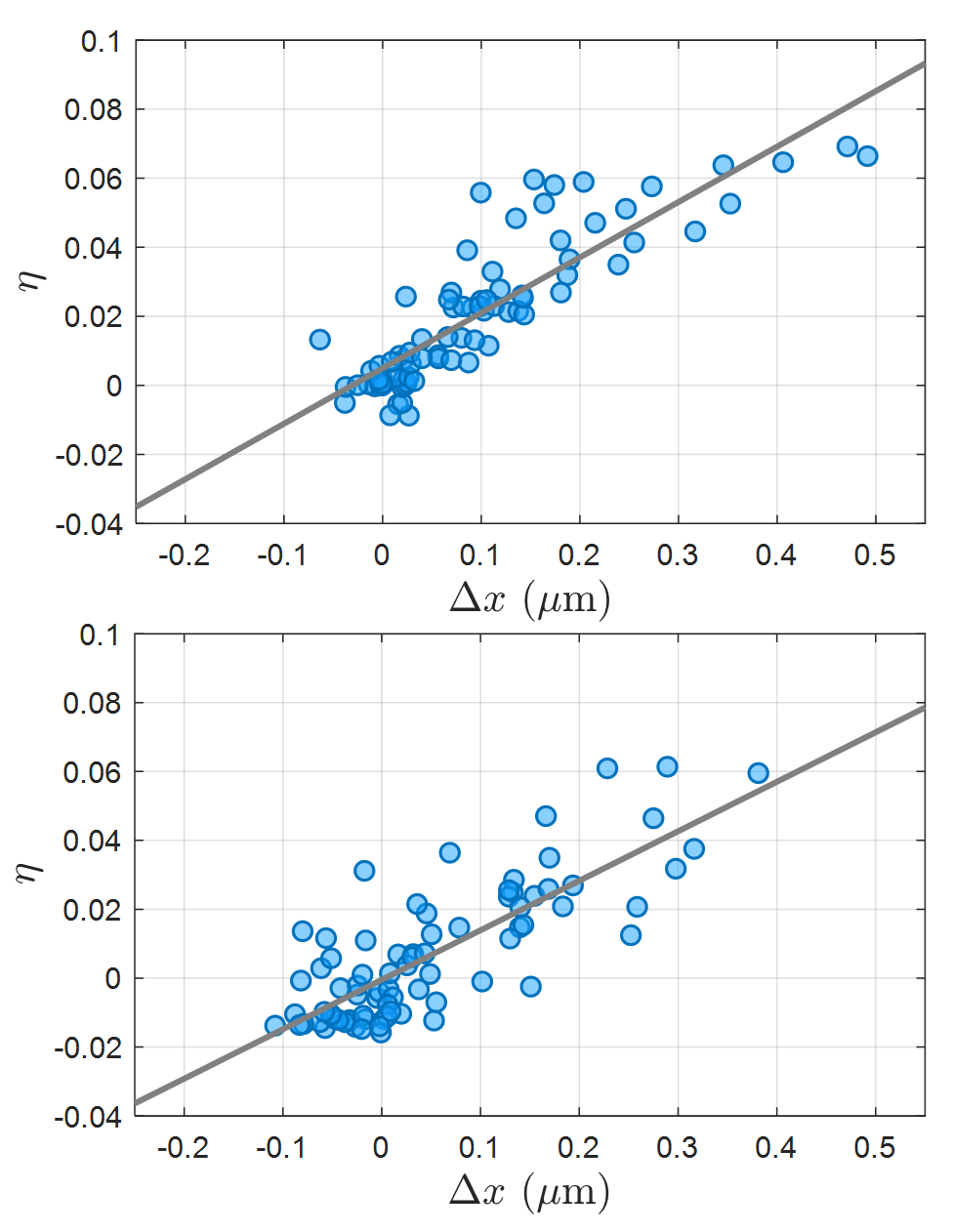}
\caption{ $\eta-\Delta x$ correlations obtained from the real time solution of Eq.~\ref{teGPE} with the prescription 1 (top) and 2 (bottom) described in the text.
 The blue circles are the numerical results for $\Delta x$ and $\eta$ at different times while the gray lines correspond to a linear fit. }
\label{figS3}
\end{figure}

In order to obtain the theory results for the $\eta-\Delta x$ correlations at $\theta=70^o$ shown
in Fig.~3 of the main text, we simulate the experimental protocol by solving 
Eq.~\ref{teGPE} while introducing a source of asymmetry in the calculations to 
force a stripe imbalance. We do so by applying two different 
prescriptions: \textit{a)} employing an initial state with a slight positive shift of
$\Delta x_0 = 1.3 \mu$m in the $x$-axis and \textit{b)} introducing an extra,
linear potential of the form $V_{\rm linear}(x) = \frac{0.01 x}{50 r_0}$ with 
$r_0 = 12 \pi a_{\rm dd}$. We then monitor the condensate wave function every 2 
ms and obtain the quantities $\eta$ and $\Delta x$ as detailed in the main text. 
We show in Fig.~\ref{figS3} the results, together with a linear fit. We can see 
that both procedures generate a quantitatively similar slope of the linear fit, 
given by $s_1 = 0.160$ $\mu$m$^{-1}$ (plotted in Fig.~3d, $s_2 =0.143$ $\mu$m$^{-1}$, which are also 
close to the value $s_{exp} = 0.155$ $\mu$m$^{-1}$ found in the experiment for $\theta=70^o$, as shown in Fig.~3
of the main text.

\subsection{Collective oscillations via the teGPE}

\begin{figure}[t]
\centering
\includegraphics[width=0.65\linewidth]{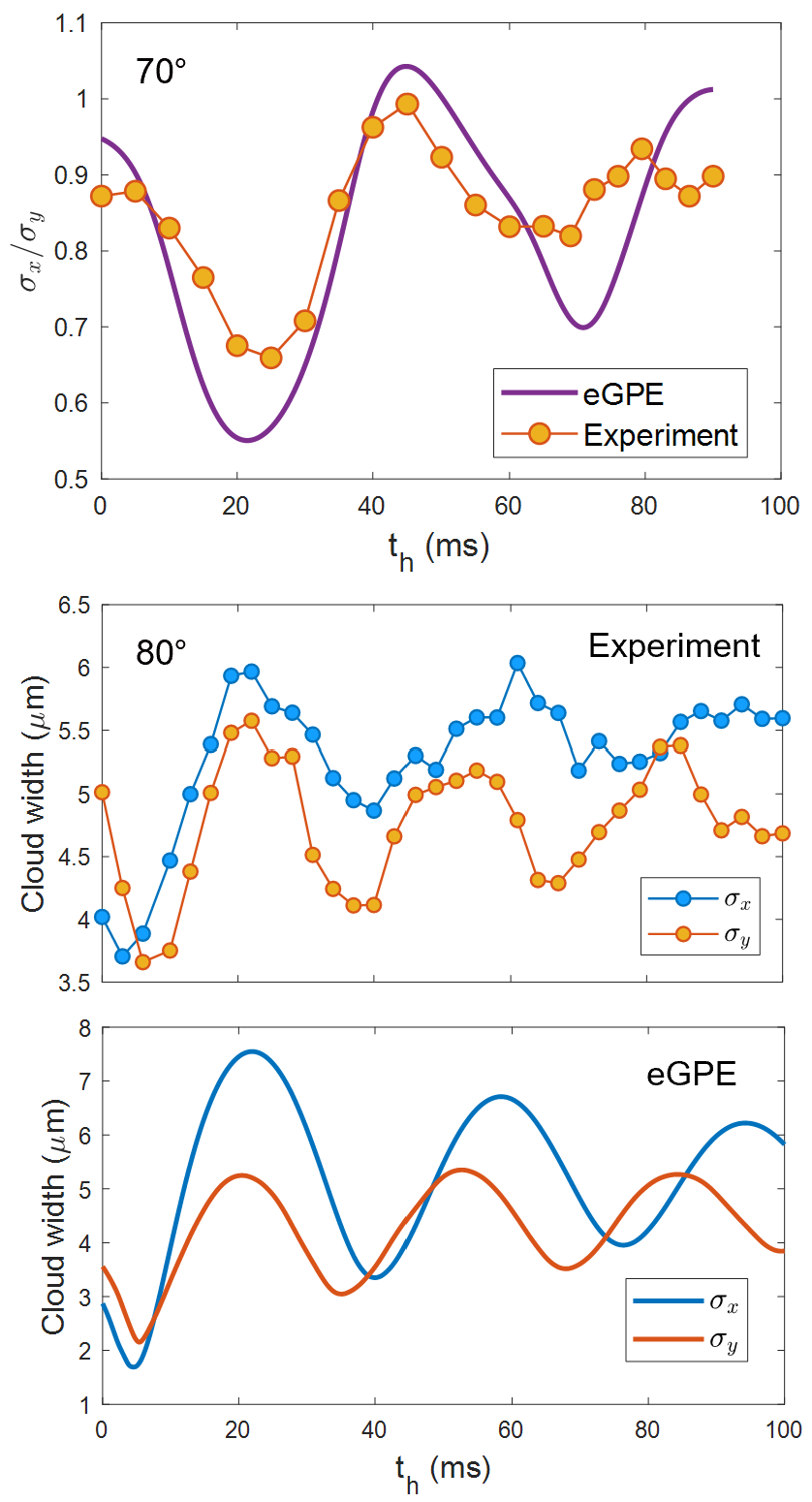}
\caption{ (Top) Aspect ratio of the widths $\sigma_x/\sigma_y$ of the atomic cloud vs holding time for $\theta=70^o$. Middle and bottom panels show the oscillations of widths $\sigma_x,y$ in experiments and simulations, respectively, vs holding time for $\theta = 80^o$.
}
\label{figS4}
\end{figure}

In order to reproduce the experimental results of Fig.~4 of the main text in
regards to the phase coherence of the experimental states, we numerically solve 
Eq.~\ref{teGPE} for $\theta=70^o$ and $\theta=80^o$ for a total time of $t=100$ 
ms and compute the widths of the wave function of the system via the expression 
$\sigma_\xi = \sqrt{ \langle \xi^2 \rangle - \langle \xi \rangle^2 }$ with $\xi 
= x,y$. We show a comparison between the theoretical and experimental results in 
Fig.~\ref{figS4}. Remarkably, we qualitatively reproduce the experimental 
oscillation of the ratio $\sigma_x/\sigma_y$ for $\theta = 70^o$, capturing the 
oscillation frequency of approximately $f^{\rm osc} \simeq \sqrt{2 f_x f_y}$ 
corresponding to the hydrodynamic limit, signaling the global phase coherence of 
the supersolid striped state. In contrast, for the case with $\theta=80^o$, the 
striped state is conformed by isolated stripes as seen in Fig.~2 of the main 
text, meaning that global phase coherence is lost. As such, it is expected that 
the width along each axis follows an oscillation with a frequency of 
approximately $f^{\rm osc}_{x,y} \simeq 2 f_{x,y}$, according to the 
collisionless regime. As it can be seen from the results, our theoretical 
calculations match the frequency of the oscillations of the experiment. This can 
be quantitatively assessed by computing the Fourier transform of both signals,
$\Sigma_{x,y}(\omega) = \int dt \sigma_{x,y}(t) e^{-i \omega t}$, for both the 
theoretical and experimental data. The main peak of $\Sigma_{x}$ for the theory 
and experimental data are located at $\omega_x^{\rm theory} = 0.157$ ms$^{-1}$ 
and $\omega_x^{\rm exp} = 0.175$ ms$^{-1}$, respectively, while for $\Sigma_{y}$ 
we have $\omega_y^{\rm theory} = 0.199$ ms$^{-1}$ and $\omega_y^{\rm exp} = 
0.215$ ms$^{-1}$. The quantitative discrepancy in the magnitude of the 
amplitudes of $\sigma_x$ and $\sigma_y$ between theory and experiment is due to 
the experimental thermal cloud, which is absent in the simulations.

\subsection{Liquid phase and gas phase}

\begin{figure}[t]
\centering
\includegraphics[width=0.65\linewidth]{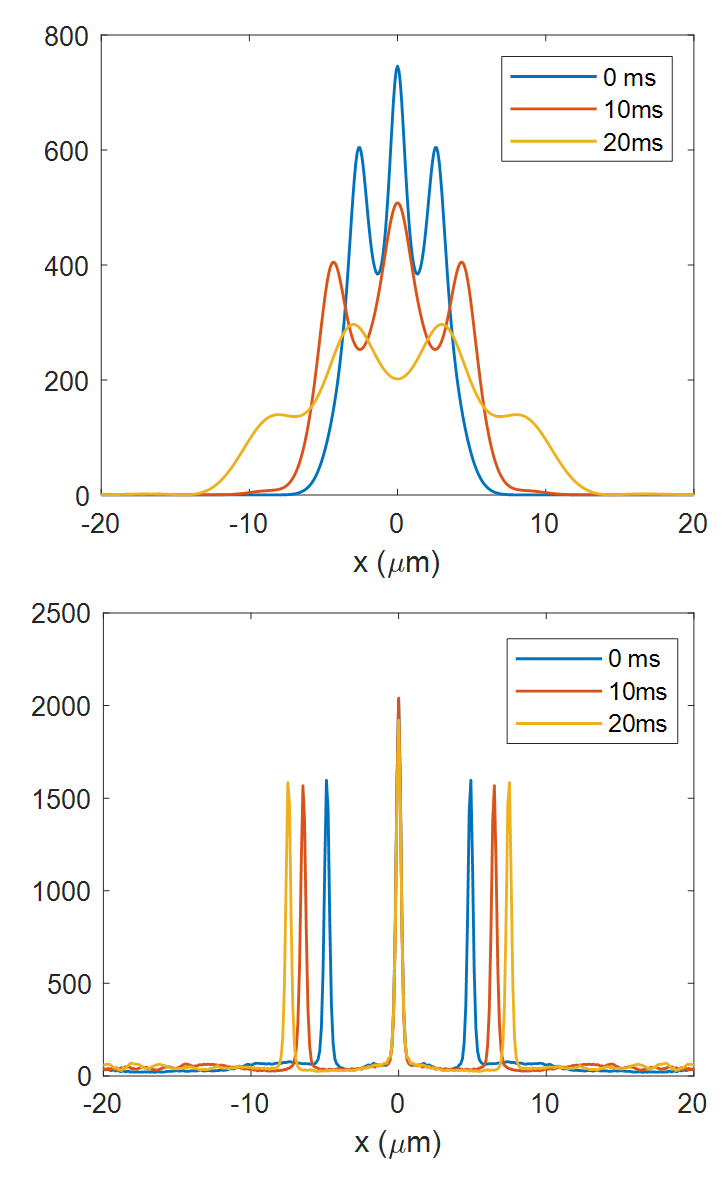}
\caption{ Column densities $\rho(x) = \int dy dz \abs{\psi(\bf r)}^2$ obtained during the 2D TOF sequence for $\theta = 70^o$ (top) and $\theta = 80^o$ (bottom).
}
\label{figS5}
\end{figure}

As reported in Ref.~\cite{sanchez2025tilted}, the phase diagram of the 
homogeneous quasi-2D dipolar system trapped along the $z$ axis features gas, 
homogeneous liquid, and stripe liquid phases at the 2D equilibrium density. 
This means that the quantum states of the trapped system can be of gas or liquid 
character. To elucidate the nature of the experimental stripe states reported
in Fig.~2 of the main text, we simulate a 2D time-of-flight expansion by solving 
Eq.~\ref{teGPE} setting $\omega_x$ and $\omega_y$ to zero, while keeping the interaction parameters $\theta$ and $\epsilon_{dd}$ constant. The initial state is 
taken as the result of the real time evolution simulating the experimental ramp 
of the scattering length followed by the corresponding holding time. We note that the TOF expansion simulated here is different from the one described in the main text and employed for matter-wave interference measurements. We show the 
results in Fig.~\ref{figS5} We see that for $\theta = 70^o$, the atomic density 
expands while the modulations melt, meaning that the stripe state is a gas. 
Instead, for $\theta=80^o$, the 2D TOF simulation shows the stripes moving 
slightly away from each other while retaining their shape. This indicates a 
stripe liquid character. In this case, each stripe is an elongated, self-bound, 
liquid droplet.

\providecommand{\noopsort}[1]{}\providecommand{\singleletter}[1]{#1}%